\documentstyle[aps,psfig,floats,epsf,twocolumn]{revtex}

\begin{document}
\tighten
\draft
\title{Anisotropic conductivity in superconducting NCCO}

\author{$^1$U. Michelucci, $^1$A.P. Kampf, and $^2$A. Pimenov}
\address{$^1$Theoretische Physik III, 
$^2$Experimentalphysik V, \\
Elektronische Korrelationen und Magnetismus \\
Institut f\"ur Physik, Universit\"at Augsburg, D-86135
Augsburg, Germany}  
\address{~
\parbox{14cm}{\rm 
\medskip
The low temperature behaviour of the in-plane and c-axis conductivity
of electron-doped cuprates like NCCO is examined; it is shown to be
consistent with an isotropic quasiparticle scattering rate and
an anisotropic interlayer hopping parameter which is non-zero for
planar momenta along the direction of the d$_{x^2-y^2}$ 
order parameter nodes. Based on these hypotheses
we find that both, the in-plane and the c-axis
conductivity, vary linearly with temperature, in agreement
with experimental data at millimiter-wave frequencies.
}}

\maketitle

\vspace{1cm}

The pairing symmetry of electron-doped cuprate materials has been the
subject of renewed interest in the last year. 
Previously existing experimental data 
were interpreted to be consistent with an s-wave order parameter
\cite{review}. In particular penetration depth measurements
\cite{Wu}, Raman scattering \cite{Barb}, and tunneling
data \cite{alff} were explained in this way.
Surprisingly, however, recent microwave experiments
\cite{Kokales,Prozorov} and phase-sensitive tricrystal
experiments \cite{tsuei} have provided evidence
in favor of a d$_{x^2-y^2}$-wave pairing symmetry.

\begin{figure}[h]
\centerline{%
\psfig{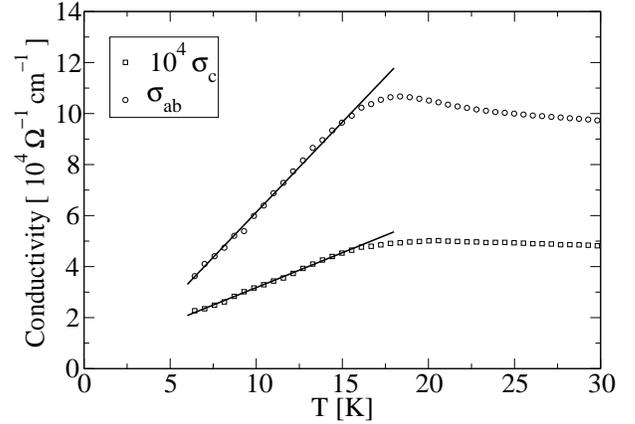}}
\caption{Experimental data for the real part of the c-axis
($\sigma_c$) and the in-plane conductivity ($\sigma_{ab}$) of
a NCCO film with $T_c=20$ K at the frequency of $\nu=8$
cm$^{-1}$ (taken from Ref. [11]). The solid lines are linear fits to the
low temperature part of the data.} 
\label{fit}
\vspace{.4cm}
\end{figure}

In this report we examine the low temperature
behaviour of the anisotropic conductivity in superconducting electron
doped cuprates. 
Using a tilted film geometry, Pimenov et al.
\cite{pimenov} have 
recently measured simultaneously $\sigma_c$ and $\sigma_{ab}$
($c$-axis and in-plane conductivity) on
Nd$_{2-x}$Ce$_x$CuO$_4$ (NCCO) samples.
By measuring the transmission through thin films as well as the
transmission phase shift with an interferometer, both complex
conductivities were determined in the submillimiter frequency range.
Their results show a characteristic temperature dependence of
$\sigma_{c}$ and $\sigma_{ab}$ in the superconducting state, varying both
linearly with $T$ almost up 
to $T_c$. Fig. \ref{fit} shows an example of these experimental
data  \cite{pimenov} for a NCCO film with $T_c\approx 20$ K. 
Fits to the low temperature data are added to underline the
linear $T$ dependence of both conductivities.  In what follows we
show how this linear $T$ dependence is consistent
with a d$_{x^2-y^2}$-wave order parameter under certain assumptions
for the scattering rate ($\Gamma_{\bf k}$) and the
c-axis hopping amplitude ($t_\perp({\bf k})$). 

For tetragonal cuprate compounds the interlayer c-axis hopping amplitude has
the form  
\cite{ander}  
\begin{equation}
t_\perp({\bf k}) = -t_\perp \cos(k_z d) (\cos(k_x a)-\cos(k_y a))^2,
\label{tperp1}
\end{equation}
where $k_x$, $k_y$ are the components of the in-plane momentum, $k_z$
the momentum along the $c$ axis and $a$ and $d$ are the corresponding
lattice constants. From now on $a$ will serve as a length unit and is
set to 1.
The form of $t_\perp({\bf k})$ in Eq. (\ref{tperp1})
results from the hybridization between the bonding oxygen 2p  and
copper 4s orbitals in each CuO$_2$ plane which gives rise to the nodal
structure of Eq. (\ref{tperp1}). 
A small deviation from the tetragonal structure leads to
a finite value of $t_\perp({\bf k})$ also along the planar nodal
directions. 
Similarly, isotropic impurity scattering is expected to effectively
contribute to the $c$-axis transport.
In this report we consider therefore the following extended form
for $t_\perp({\bf k})$  
\begin{equation}
t_\perp({\bf k}) = -t_\perp [\cos(k_zd)(\cos(k_x)-\cos(k_y))^2+c].
\label{tperp}
\end{equation}

We are interested in the real part of the dc limit of the longitudinal
conductivity both for the in-plane and the c-axis case.
The conductivity, labeled by
$\sigma_{\mu\mu}$, where $\mu$ is either $z$ or $x$, is given by:
\begin{equation}
\sigma_{\mu\mu}' \equiv \hbox{Re} \ \sigma_{\mu \mu} =\lim_{\omega
\rightarrow 0} 
\frac{\hbox{Im} \ D_{\mu \mu} (\omega+i0^+)}{\hbar \omega}
\label{eq3}
\end{equation}
where $D_{\mu \mu} (i\Omega)$ is the current-current correlation function
\begin{equation}
D_{\mu \mu} (i\Omega) = \lim_{{\bf q} \rightarrow {\bf 0}}
\int_0^\beta e^{i\Omega 
\tau} \langle T_\tau \hat \jmath_\mu ({\bf q},\tau) \hat
\jmath_{\mu}({\bf -q},0) \rangle d\tau
\label{dmu}
\end{equation}
with $\hat \jmath_\mu({\bf q},\tau)$ the paramagnetic current density
operator. Note that the usual notation $\sigma_c$ corresponds to
$\sigma_{zz}$ and 
$\sigma_{ab}$ to $\sigma_{xx}$.
In Eq. (\ref{dmu}) $\beta=1/T$, $T$ the temperature, $T_\tau$ the
imaginary time ordering operator and $\Omega$ is a bosonic Matsubara
frequency.  
For the purpose of comparing with experimental data in the
submillimeter frequency range presented in Fig. \ref{fit} the zero
frequency limit in Eq. (\ref{eq3}) is appropriate.

Neglecting vertex corrections we rewrite $D_{\mu \mu} (i\Omega)$ as
\begin{equation}
D_{\mu \mu}(i\Omega) = \frac{1}{\beta N} \sum_{{\bf k},i\omega}
j_\mu^2({\bf k}) 
\hbox{Tr} \{ G({\bf k},i\omega) G({\bf k},i\omega+i\Omega) \}. 
\end{equation}
$j_{x(z)}({\bf k}) = e/\hbar (\partial \xi_{\bf k} / \partial k_{x(z)})$
is the current vertex function and $G({\bf 
k},i\omega)$ is the renormalized BCS Green's function in Nambu
space 
\begin{equation}
G({\bf k},i\omega) = \frac{(i\tilde \omega)\tau_0 +\tilde \xi_{\bf
k}\tau_3 +\tilde \Delta_{\bf k}\tau_1}{(i\tilde \omega)^2- \tilde
\xi_{\bf k}^2- \tilde \Delta_{\bf k}^2}
\end{equation}
where $\tilde \omega = \omega - \Sigma_0$, $\tilde \xi_{\bf k} =
\xi_{\bf k} + \Sigma_3$ and $\tilde \Delta_{\bf k}=\Delta_{\bf k}
+\Sigma_1 $.
$\xi_{\bf k}$ is the band dispersion measured with respect to the
chemical potential, $\xi_{\bf k}=\epsilon_{\bf k}-\mu$, $\tau_i$ are
the Pauli matrices, and $\Delta_{\bf k}$ is the superconducting gap
that we take in a d$_{x^2-y^2}$-wave form 
\begin{equation}
\Delta_{\bf k} = \frac{\Delta_0}{2} (\cos(k_x)-\cos(k_y)).
\end{equation}
$\Sigma_i$ are the components of the self energy matrix $\Sigma=\sum_{i=0}^3
\Sigma_i \tau_i$ assumed to dominantly arise from impurity scattering
in the low frequency and low temperature limit that we consider here.
Using the spectral representation for $G$ and summing over the
internal Matsubara frequencies the real part of the dc
conductivity results as
\begin{equation}
\begin{array}{lll}
\medskip \sigma'_{\mu \mu} &=& \displaystyle \frac{\hbar}{N}
\sum_{\bf k} j_\mu^2({\bf k}) \int d\epsilon \ \frac{\partial
f(\epsilon/T)}{\partial \epsilon} \hbox{Tr} \{\hbox{Im} \ G({\bf
k},\epsilon) \times \\
&&\hbox{Im} \ G({\bf k},\epsilon) \} 
\end{array}
\label{eq8}
\end{equation}
where $f(x)=(\exp(x)+1)^{-1}$ is the Fermi function.
In the following we assume that the band dispersion $\xi_{\bf k}$
depends only on $k_z$ and $k=\sqrt{k_x^2+k_y^2}$ and we evaluate the
momentum sum as 
\begin{equation}
\frac{1}{N} \sum_{\bf k} \rightarrow N(0) \int_{-\pi}^{\pi} d k_z
\int_0^{2\pi} d\phi \int d\tilde \xi_{\bf k}.
\end{equation}
$\phi$ is the azimuthal angle around the cylindrical Fermi surface
(FS) and the density of states is approximated by its value at the FS 
$N(\tilde \xi_{\bf k}) \approx N(0)$. 
All the functions in Eq. (\ref{eq8}) are then parametrized along the FS
as functions of $\phi$, $k_z$ and $\tilde \xi_{\bf k}$.
Note that in what follows we neglect the effect of $\Sigma_1$,
i.e. any gap renormalization, since
we are interested in the effect of the nodal structure of a pure
d$_{x^2-y^2}$-wave gap.
For the purpose of calculating the dc conductivity limit it is
furthermore justified to ignore $\hbox{Im} \ \Sigma_3$ which is
negligibly small in the very low frequency limit \cite{parks}.
We
set $\Sigma_0=-i\Gamma$ where $\Gamma$ is an
isotropic impurity scattering rate on the FS. We note that an 
anisotropic form for $\Sigma_0$ leads to higher order terms in the
subsequent low temperature expansion \cite{xiang} but leaves
unaffected the leading low temperature behaviour.

With a d-wave gap
$\Delta_\phi=\Delta_0 \cos(2 \phi)$ and $t_\perp(\phi) = -t_\perp
[\cos(k_zd) \cos^2 2\phi + c]$
the result after integration over $\tilde \xi_{\bf k}$
and $k_z$ is written in the form:
\begin{equation}
\begin{array}{lll}
\medskip \sigma_{\mu \mu}' & = & -\alpha_{\mu \mu} \displaystyle
\int_{-\infty}^{\infty} 
d\epsilon \ \frac{\partial f(\epsilon/T)}{\partial \epsilon}
 \int_0^{2\pi} d\phi \ u_{\mu \mu}^2(\phi)\cdot \\
&&\hbox{Re} \displaystyle \frac{(\epsilon+i\Gamma)^3-\epsilon
 \Delta_\phi^2}{\Gamma [(\epsilon+i\Gamma)^2
 -\Delta_\phi^2]^{3/2}}    
\end{array}
\label{eq10}
\end{equation}
Here we have defined $\alpha_{xx}=2\hbar e^2N(0)v_F^2$ with $v_F$ the
Fermi velocity in the 
$ab$ plane, $\alpha_{zz}=e^2N(0)t_\perp^2d^2/\hbar$, $u_{xx}(\phi) =
1$ and $u_{zz}(\phi) = (\cos^2 (2\phi)+c)$.

The energy integral in Eq. (\ref{eq10}) can be divided in two
parts, the first with $|\epsilon| \in [0,\Gamma]$ and the second with
$|\epsilon| \in [\Gamma,\infty]$.
To analyze these two contributions separately we introduce
the following notation
\begin{equation}
\sigma'_{\mu \mu} = \int^{\infty}_{-\infty} d\epsilon \ \frac{\partial
f(\epsilon/T)}{\partial \epsilon} \Lambda_{\mu \mu} (\epsilon)
\label{eq11}
\end{equation}
where, comparing Eq. (\ref{eq10}) and Eq. (\ref{eq11}),
\begin{equation}
\Lambda_{\mu \mu}(\epsilon) = -\alpha_{\mu \mu} \int_0^{2\pi} d\phi \ u_{\mu
\mu}^2(\phi) 
\hbox{Re} \displaystyle \frac{(\epsilon+i\Gamma)^3-\epsilon 
 \Delta_\phi^2}{\Gamma [(\epsilon+i\Gamma)^2
 -\Delta_\phi^2]^{3/2}}.
\end{equation}
Eq. (\ref{eq10}) is thus rewritten as
\begin{equation}
\begin{array}{lll}
\medskip \sigma_\mu' & = & \displaystyle \int_{-\infty}^{\infty}
d\epsilon \ \frac{\partial f(\epsilon/T)}{\partial \epsilon}
 \Lambda_{\mu \mu}(\epsilon) \Theta(-|\epsilon|+\Gamma)+\\
&&\displaystyle \int_{-\infty}^{\infty}
d\epsilon \ \frac{\partial f(\epsilon/T)}{\partial \epsilon}
 \Lambda_{\mu \mu}(\epsilon) \Theta(|\epsilon|-\Gamma)
\end{array}
\label{eq12}
\end{equation}
where $\Theta(x)$ is the step function.
Since we are interested in the low temperatures behaviour we perform
a Sommerfeld like expansion of the first integral, leading to a dominant
$T$ independent contribution $\sigma_{0,\mu \mu}'$. 
The second integral is then evaluated using an expansion in
$\epsilon/\Gamma$ for the function
$\Lambda_{\mu \mu}(\epsilon)$ \cite{ioffe}.

The resulting leading terms for the longitudinal conductivity in powers of
$T/\Delta_0$ are given by
\begin{equation}
\begin{array}{l}
\bigskip \sigma_{zz}' = \displaystyle  
  \sigma_{0,zz}' + \frac{e^2}{\hbar} \frac{4\pi (\ln 2) N(0)
  t_\perp^2 d^2 c^2}{\Gamma}
  (T/\Delta_0) +O(T^2) 
\\ 
\sigma_{xx}'= \displaystyle  
  \sigma_{0,xx}'+\frac{e^2}{\hbar} \frac{8\pi (\ln 2) N(0)
  v_F^2 \hbar}{\Gamma} (T/\Delta_0) +O(T^2).
\end{array}
\label{lineart}
\end{equation}
These expansions are valid in the limit $\Gamma < \Delta_0$
expected to apply for NCCO \cite{pimenov}.

$\sigma_{0,\mu \mu}'$ are the
limits $\omega \to 0$, $T\to 0$ of the conductivities. Explicitly they
are given by
\begin{equation}
\begin{array}{l}
\medskip \sigma_{zz,0}' = \displaystyle \frac{4 e^2N(0)t_\perp^2 d^2
c^2}{\Delta_0^2 \hbar} \\
\sigma_{xx,0}' = \displaystyle \frac{8 e^2 N(0) v_F^2 \hbar}{\Delta_0}.
\end{array}
\label{universal}
\end{equation}
From Eq. (\ref{universal}) we note that $\sigma'_{xx,0}$ is
independent of impurity controlled parameters such as $\Gamma$ and possibly
$c$. This fact is consistent with the work by Lee \cite{Lee}, in which
a universal limit of the optical conductivity for $\omega\to 0$
and $T \to 0$ is found, i.e. independent of impurity scattering.

The linear temperature dependence of $\sigma_{zz}'$ and $\sigma_{xx}'$ in
Eq. (\ref{lineart}) is clearly consistent with the experimental data on
NCCO \cite{pimenov}. In addition we may ask whether Eq. (\ref{lineart}) is
also quantitatively 
consistent with the data. We have argued that $c$ originates from
small effects like deviations from the ideal tetragonal crystal
structure or isotropic impurity scattering, and so we 
expect $t_\perp(\phi=\pi/4) \ll t_\perp(\phi=0)$, or equivalently $c
\ll 1$. 
We estimate $c$ from the ratio $r$ of the slopes of the two
conductivities $\sigma_{zz}'$ and $\sigma_{xx}'$, i.e.
\begin{equation}
r \equiv \frac{\sigma_{zz}'-\sigma_{zz,0}'}{\sigma_{xx}'-\sigma_{xx,0}'} =
\frac{t_\perp^2 c^2}{2 \hbar^2 v_F^2} d^2.
\end{equation}
From the data on NCCO shown in Fig. \ref{fit} we find $r
\approx 3 \cdot 10^{-4}$. Using  $d \sim 3\cdot 10^{-10}$ m, $v_F 
\approx 2.2 \cdot 10^5$ m/s \cite{homes}, and $t_\perp \approx 10$
meV the resulting estimate for $c$ is 
\begin{equation}
c \sim 10^{-2}
\end{equation}
verifying our original assumptions.

We point out that the result obtained in Eq. (\ref{lineart}) is
due to the d$_{x^2-y^2}$-wave form for the superconducting
gap combined with the modified $c$-axis hopping amplitude Eq (\ref{tperp}). A
different symmetry for the order parameter leads to a very different low
temperature behaviour for the conductivities. 
To this extent it is useful to compare these results with those
calculated for an s-wave gap.
Substituting $\Delta_\phi=\Delta_0 \equiv \hbox{const.}$ in
Eq. (\ref{eq10}) the low 
temperature expansion is now given by
\begin{equation}
\begin{array}{l}
\sigma_{zz}^s = 
\displaystyle 
  \sigma_{0,zz}^s+ \frac{e^2}{\hbar} c_{zz} e^{-\Delta_0/T} \left(
  \sqrt{\Delta_0/T} +O(\sqrt{T/\Delta_0}) \right)
\\
\sigma_{xx}^s =
\displaystyle 
  \sigma_{0,xx}^s + \frac{e^2}{\hbar} c_{xx}e^{-\Delta_0/T} \left(
  \sqrt{\Delta_0/T} +O(\sqrt{T/\Delta_0}) \right)
\end{array}
\end{equation}
where $c_{zz}$ and $c_{xx}$ are two constants.
This result is clearly incompatible with the experimental data.

A similar analysis as presented above has been previously applied to
cuprates superconductors like YBCO 
\cite{xiang,ioffe} with a different form for the
interlayer hopping amplitude, namely Eq. (\ref{tperp}) has been used with
$c=0$. This leads to a different low temperature behaviour
for $\sigma_c$ \cite{xiang}, namely $T^3$, that was shown to be compatible
with experimental data on optimally doped YBCO and BSCCO\cite{xiang}. It is
important to note that the only 
difference between our assumptions and the ones used for YBCO is the
presence of the constant $c$ in $t_\perp({\bf k})$ that allows us to
obtain a low temperature behaviour of $\sigma$ compatible with
NCCO data.

In conclusion we have shown how, with an isotropic scattering rate and
an interlayer hopping integral non-zero along the Brillouin zone diagonals,
it is possible to obtain a linear T dependence for the low temperature
behaviour of the conductivities at millimeter wavelengths.
This is indeed observed experimentally and supports a
d$_{x^2-y^2}$-wave pairing symmetry for electron-doped materials like
NCCO. 

\vspace{.2cm}

This work was partially supported by the Deutsche Forschungsgemeinschaft
through SFB 484.

\end{document}